\documentclass[journal]{IEEEtran}

\usepackage{cite}
\usepackage{graphicx}
\usepackage{amsmath}
\usepackage{amssymb}
\usepackage{amsfonts}
\usepackage{array}
\usepackage{mathtools}
\usepackage[ruled,vlined,linesnumbered]{algorithm2e}
\usepackage{algorithmic} 
\usepackage{url}
\usepackage{float}

\usepackage[table,svgnames,HTML]{xcolor}
\definecolor{green}{HTML}{1A8033}
\definecolor{blue}{HTML}{4DB3E6}
\definecolor{red}{HTML}{E64D4D}

\usepackage{tabularx}
\usepackage{multirow}
\usepackage{makecell}
\usepackage{booktabs}

\usepackage{setspace}

\newcolumntype{P}[1]{>{\raggedright\arraybackslash}p{#1}} 

\usepackage[hidelinks]{hyperref}

\begin{document}

\title{Large Wireless Foundation Models:\\ Stronger over Bigger}

\author{
Xiang Cheng,~\IEEEmembership{Fellow,~IEEE}, Boxun Liu,~\IEEEmembership{Graduate Student Member,~IEEE}, \\Xuanyu Liu,~\IEEEmembership{Graduate Student Member,~IEEE}, and Xuesong Cai,~\IEEEmembership{Senior Member,~IEEE}
\thanks{X. Cheng, B. Liu, X. Liu, and X. Cai are with the State Key Laboratory of Photonics and Communications, School of Electronics, Peking University, Beijing 100871, China (email: \url{xiangcheng@pku.edu.cn};\url{boxunliu@stu.pku.edu.cn};\url{xyliu25@stu.pku.edu.cn};\url{xuesong.cai@pku.edu.cn})}
}

\markboth{Journal of \LaTeX\ Class Files,~Vol.~X, No.~X, January~2026}%
{Journal of \LaTeX\ Class Files,~Vol.~X, No.~X, January~2026}

\maketitle

\begin{abstract}
AI–communication integration is widely regarded as a core enabling technology for 6G. 
Most existing AI-based physical-layer designs rely on task-specific models that are separately tailored to individual modules, resulting in poor generalization.
In contrast, communication systems are inherently general-purpose and should support broad applicability and robustness across diverse scenarios.
Foundation models offer a promising solution through strong reasoning and generalization, yet wireless-system constraints hinder a direct transfer of large language model (LLM)-style success to the wireless domain. 
Therefore, we introduce the concept of large wireless foundation models (LWFMs) and present a novel framework for empowering the physical layer with foundation models under wireless constraints. 
Specifically, we propose two paradigms for realizing LWFMs, including leveraging existing general-purpose foundation models and building novel wireless foundation models. 
Based on recent progress, we distill two roadmaps for each paradigm and formulate design principles under wireless constraints. We further provide case studies of LWFM-empowered wireless systems to intuitively validate their advantages. 
Finally, we characterize the notion of “large” in LWFMs through a multidimensional analysis of existing work and outline promising directions for future research.
\end{abstract}

\section{Introduction}
As a new generation of critical infrastructure, the sixth-generation mobile communication network (6G) is expected meet the broader and more complex demands of emerging services, including cloud virtual reality (VR), cellular vehicle-to-everything (C-V2X), and digital twins. 
Artificial intelligence (AI), notably deep learning, offers powerful data-driven modeling capability, enabling it to learn complex mapping relationships directly from raw data without prior assumptions, and thus serving as an important complement to conventional model-based wireless algorithms.
Therefore, the integration of AI and communications \cite{SoM} is widely acknowledged as a core enabling technology of 6G, and is expected to profoundly reshape design paradigms across network layers, particularly at the physical layer.

Over the past decade, AI has been extensively explored by both academia and industry for wireless physical-layer design, covering diverse modules such as channel state information (CSI) feedback, channel estimation and prediction, and equalization. 
Compared with conventional parametric model-based physical-layer algorithms, AI-based approaches can learn an implicit representation of complex real-world propagation environments from data, thereby substantially improving the spectral efficiency and robustness. 
Nevertheless, most existing AI-based physical-layer solutions remain dominated by task-specific models, contradicting the generality of communication systems.
Specifically, they are tailored to individual modules and system configurations and are trained for specific channel distributions, leading to the following evident limitations.

\begin{itemize}
    \item \textbf{Insufficient capability}: Existing studies are dominated by small deep-learning models, whose limited scale makes it difficult to handle highly dynamic environments and complex physical-layer tasks.
    \item \textbf{Limited generalization}: When the wireless data distribution shifts, task-specific AI models must be retrained, incurring additional costs in wireless data collection and training.
    \item \textbf{Fragmented design}: Task-specific AI models are developed separately for individual physical-layer modules, which sharply increases storage, computation, and model management overhead.
\end{itemize}

In recent years, the rise of foundation models has reshaped the deep-learning paradigm and has revolutionized multiple fields such as natural language processing (NLP) and computer vision (CV), spawning large language models (LLMs), large vision models (LVMs), and multimodal large language models (MLLMs) exemplified by ChatGPT and DeepSeek.
Through large-scale pretraining, foundation models achieve reasoning and generalization beyond task-specific AI models, opening new avenues for tackling these challenges and enabling general-purpose wireless AI.
However, LLMs, as their name suggests, are characterized by large parameter counts, massive training data, and high computational cost, which fundamentally conflict with the stringent constraints of wireless communication systems, including the power and memory limits of communication devices, as well as the latency and reliability requirements inherent to communication tasks. 
This raises a fundamental question of \textit{whether and how to build large wireless foundation models (LWFMs) under stringent communication constraints}.

Existing studies have explored the application of foundation models to wireless physical layer design.
However, a systematic methodological synthesis remains lacking.
Our previous survey \cite{tnsewifo} is the first to introduce two paradigms for the foundation models empowered synesthesia of machines (SoM) system design, while offering a limited in-depth discussion on how foundation models can empower the wireless domain under stringent communication constraints.
The most recent studies \cite{LAM4PHY} have attempted to review related works under the umbrella of large AI models \cite{BigAI6G}, yet they lack a precise definition and systematic analysis of what “large” entails. 

To fill these gaps, we propose the concept of LWFM and systematically develop it as a new framework for wireless system design empowered by foundation models.
We first explain the motivation and discuss the challenges of applying large models to wireless systems. 
We then present two paradigms of LWFM, including leveraging general-purpose foundation models and building wireless foundation models. 
For each paradigm, we describe the underlying concept, provide representative roadmaps, distill design principles under wireless constraints, and summarize the key features. 
We further include a case study to provide an intuitive illustration of the advantages of LWFM in wireless systems.
Finally, we conclude and outline promising directions for future research.

\begin{table*}[!ht]
\centering
\footnotesize
\caption{Comparison of constraints of wireless communication systems and requirements of LLMs.}
\label{comparison}
\renewcommand{\arraystretch}{1.15}
\begin{tabularx}{\textwidth}{|>{\centering\arraybackslash}m{2cm}|>{\centering\arraybackslash}m{2cm}|m{7.6cm}|m{4.8cm}|}
\hline
\multicolumn{2}{|c|}{\textbf{Constraint and requirement type}} &
\textbf{Constraints of wireless communication systems} &
\textbf{Requirements of LLMs} \\
\hline
\multirow{2}{2cm}{\centering Hardware\\ level}
& Compute
& The distributed unit (DU) of the base station (BS)  is typically equipped with only SoC (system on a chip) or FPGA (field-programmable gate array) offering on the order of \textbf{100 TOPS} of compute, and is generally not provisioned with a GPU.
Advanced smartphones typically come with a GPU delivering less than \textbf{10 TOPS} of compute.
& Server clusters used to deploy LLMs are typically equipped with multiple high-performance GPUs (e.g., NVIDIA H100), providing over \textbf{10,000 TOPS} of INT8 compute. \\
\cline{2-4}
& Storage
& The BSs typically have external memory capacity on the order of \textbf{GB}, whereas on-chip storage is generally limited to only \textbf{tens of MB}.
Smartphones typically use a unified memory architecture shared with the CPU, with total memory capacity generally on the order of around \textbf{10 GB}.
& The VRAM of a single high-performance GPU used for deploying LLMs is typically close to \textbf{100 GB}. \\
\hline
\multirow{2}{2cm}{\centering System \\ level}
& Latency
& The latency requirement of the physical-layer modules is as low as the \textbf{1 ms} level, and they must strictly satisfy timing relationships.
& The output latency of LLMs can be as long as \textbf{several minutes}, with no strict time limit. \\
\cline{2-4}

& Reliability
& In the ultra-reliable low-latency communications (uRLLC) scenario of 5G NR, the successful delivery probability for small packets needs to reach \textbf{99.999\%}.
& The outputs of LLMs exhibit inherent randomness and hallucinations. \\
\hline
\end{tabularx}
\end{table*}
\section{Large Wireless Foundation Models}
In this section, we first introduce the motivation for developing LWFMs and then describe the unique challenges they face in wireless communication systems compared with LLMs.
\subsection{Motivation}
A foundation model is pretrained on large-scale datasets in a self-supervised manner, and can be adapted to a wide range of downstream tasks through fine-tuning or zero-shot inference, often substantially outperforming task-specific models. 
Following scaling laws, its capability strengthens as the number of parameters, the amount of training data, and the available compute increase. 
In addition to NLP and CV, foundation models have also been successfully applied to domains such as biomedicine, weather forecasting, and remote sensing, validating their potential in specialized fields. 
Analogous to LLMs, domain-specific foundation models for wireless systems, namely \textbf{\textit{Large Wireless Foundation Models}}, need to be developed to realize the following potential advantages.
\begin{itemize}
    \item \textbf{Advanced modeling capability}: Foundation models can efficiently capture complex channel patterns and solve hard optimization problems, improving system throughput and robustness in harsh conditions such as highly dynamic environments and strong interference.
    \item \textbf{Emergent generalization capability}: With no fine-tuning or only a small number of samples, foundation models can perform well on previously unseen scenarios and system configurations, substantially reducing the cost of data collection and fine-tuning.
    \item \textbf{Multi-task versatility}: Foundation models can enable multiple physical-layer tasks with a single model, significantly reducing the storage and compute resources required for model deployment on communication devices.
\end{itemize}

\subsection{Challenges}
Nevertheless, unlike LLMs, LWFMs are deployed within wireless communication systems and must operate under stringent hardware and system constraints.
As shown in Table \ref{comparison}, at both the hardware and system levels, the requirements of LLMs naturally conflict with the constraints of wireless communication systems, as detailed below.
\begin{itemize}
    \item \textbf{Compute}: Computing capability determines inference speed and the level of concurrency that can be supported. LLMs are typically deployed on high-performance server clusters offering on the order of 10,000 tera operations per second (TOPS). However, constrained by power consumption and cost, communication devices are generally compute-limited. 
    BSs are commonly equipped with SoCs providing roughly 100 TOPS, and only a few advanced smartphones integrate graphics processing units (GPUs), with compute capability typically below 10 TOPS.
    \item \textbf{Storage}: Storage determines the scale of models and data that can be loaded. Deploying LLMs typically requires high-performance GPUs with around 100 gigabytes (GB) of video random access memory (VRAM). By contrast, the memory capacity of base stations and smartphones is usually on the order of a few gigabytes, and their cache size is only on the order of 10 megabytes (MB).
    \item \textbf{Latency}: Latency reflects the time constraint for completing a task. LLM inference is typically subject to relatively loose latency requirements, often on the order of minutes. In contrast, wireless physical-layer tasks must strictly adhere to timing logic and usually need to be completed on the order of milliseconds.
    \item \textbf{Reliability}: Reliability measures the accuracy of the output. LLM outputs can be highly stochastic and may exhibit hallucinations, and they may not always be consistent with reality. In contrast, wireless communication systems require extremely high reliability, often exceeding 99.999\%.
\end{itemize}

In summary, directly building LWFMs by drawing an analogy to LLMs faces multiple challenges. 
Therefore, the design paradigms for LWFMs that explicitly account for the wireless constraints need to be systematically articulated.

\begin{figure*}[!t]
    \centering
    \includegraphics[width=1\linewidth]{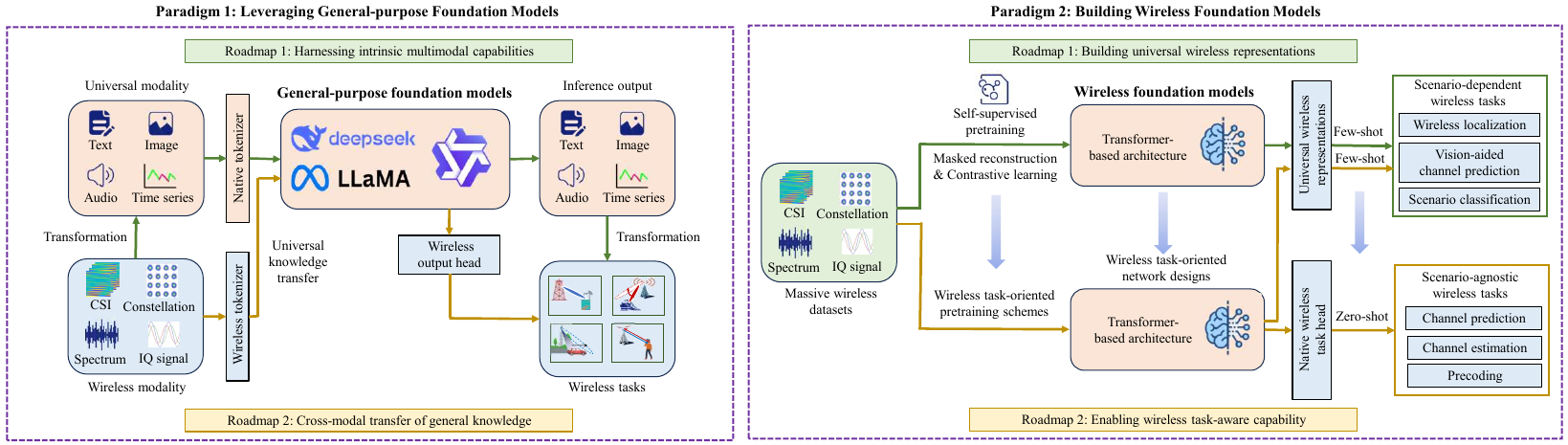}
    \caption{An illustration of the proposed two paradigms for LWFM.}
    \label{paradigm-fig}
\end{figure*}
\section{Paradigm 1: Leveraging General-purpose Foundation Models}
In the following two sections, we introduce the two proposed paradigms for realizing LWFM.
In this section, we focus on the first paradigm, which leverages general-purpose foundation models. 
We first introduce the concept of general-purpose foundation models and the motivation. 
Then we describe two representative implementation roadmaps and discuss their wireless constraint-aware designs. 
Finally, we summarize the key features of this paradigm.

\subsection{Concept}
In contrast to domain-specific foundation models, general-purpose foundation models are designed for non-specialized domains, such as LLMs, LVMs, MLLMs, and time-series foundation models. 
Pretrained on massive datasets, general-purpose foundation models, especially LLMs, have shown emergent general reasoning and generalization capabilities, providing a straightforward paradigm for building LWFMs. 
On the one hand, their strong reasoning ability in language and vision tasks can be leveraged to address analogous problems in wireless communications effectively. 
On the other hand, through cross-domain adaptation of broad semantic knowledge, general-purpose foundation models can rapidly accommodate diverse and evolving wireless communication scenarios while maintaining strong generalization.
\subsection{Representative Roadmaps}
According to whether the wireless modality is converted into modalities natively supported by general-purpose foundation models, we propose two representative implementation roadmaps for paradigm 1, as shown in Fig \ref{paradigm-fig}. 
\subsubsection{Harnessing intrinsic multimodal capabilities}
The first roadmap exploits the native multimodal capabilities of general-purpose foundation models to address wireless tasks. 
For example, LVM4CSI \cite{LVM4CSI} leverages the similarity between CSI and visual data and directly applies LVMs to channel estimation, human activity recognition, and user localization, achieving substantial gains in both accuracy and generalization. 
In addition, a recent work \cite{PFMCE} decomposes high-dimensional channel prediction into a set of parallel one-dimensional time-series forecasting problems and fine-tunes the pretrained time-series foundation model TimeFM on CSI datasets, thereby enabling strongly generalizable channel estimation under sparser pilot configurations.

\subsubsection{Cross-modal transfer of general knowledge}
Although the first roadmap preserves the ability of general-purpose foundation models to process their native modalities, its capacity to handle a broad range of other wireless modalities, such as channel state information (CSI), is limited. 
Therefore, an alternative roadmap aims to integrate communication modalities directly into general-purpose foundation models through techniques such as fine-tuning, thereby enabling the cross-modal transfer of general knowledge to wireless tasks \cite{BP_LLM}. 
As a pioneering effort, LLM4CP \cite{LLM4CP} was the first to apply an LLM to non-linguistic physical-layer tasks by fine-tuning GPT-2, achieving improved accuracy and generalization for channel prediction. 
Building on this direction, LLM4WM \cite{LLM4WM} extends the roadmap to wireless multi-tasks. It adopts a mixture of experts with low-rank adaptation (MoE-LoRA) to realize soft parameter sharing, mitigating conflicts across physical-layer tasks and delivering joint multi-task performance gains.

\subsection{Design Principles under Wireless Constraints}
For paradigm 1, we summarize the following design principles under wireless constraints.

\begin{itemize}
    \item For roadmap 1, leveraging the native modality-processing capability of general-purpose foundation models typically requires using the complete model. 
    Therefore, model selection should favor smaller general-purpose foundation models, and these models are better suited for deployment on more compute-capable communication equipment like BS.
    In addition, task selection should prioritize latency-tolerant applications, such as precoding and user localization.
    \item For roadmap 2, it is possible to more flexibly adopt only a subset of a large model rather than the complete model, thereby reducing the compute and memory requirements on communication devices. 
    Consequently, this roadmap can support more real-time tasks, such as channel prediction. 
    For example, LLM4CP adopts a 6-layer GPT-2 and achieves an inference latency comparable to that of conventional small models.
\end{itemize}
\subsection{Key Features}
The advantages and limitations of paradigm 1 are summarized as follows.
\begin{itemize}
    \item By standing on the shoulders of giants, paradigm 1 can fully leverage state-of-the-art general-purpose foundation models developed for different modalities, while avoiding the compute-intensive pretraining process. 
    Notably, the rapidly growing body of recent advances around LLMs can be seamlessly incorporated, enabling continuous improvement of wireless solutions within this paradigm. For example, research on LLM model compression and hardware acceleration is directly applicable to LLM-based wireless designs and can effectively mitigate deployment challenges under resource constraints.
    \item Nevertheless, since these models are not designed specifically for the wireless domain, their reasoning and generalization capabilities for wireless tasks remain limited. 
    For example, although LVMs can treat CSI as an image for processing, the structured multi-path characteristics of CSI are still not explicitly captured, which can lead to accuracy degradation. 
    Moreover, while existing studies show that fine-tuning LLMs can yield gains on wireless tasks, the gap between the language and wireless domains is still difficult to bridge effectively.
\end{itemize}

\section{Paradigm 2: Building Wireless Foundation Models}
In this section, we illustrate the second paradigm of LWFMs, which is to build wireless domain-specific foundation models. 
We first introduce the concept of wireless foundation models and their advantages. 
We then describe two representative roadmaps and their wireless constraint-aware designs. 
Finally, we summarize the key features of this paradigm.
\subsection{Concept}
\textbf{\textit{Wireless foundation models}} refer to domain-specific foundation models that are pretrained from scratch for the wireless domain. 
Through pretraining on massive and heterogeneous wireless datasets, they can be applied to a broad range of physical-layer tasks via few-shot or even zero-shot learning and can substantially outperform task-specific AI models. 
They are expected to serve as a unified wireless model across various physical-layer modules, thereby markedly reducing the number of AI models required in practice. 
In addition, they can generalize across heterogeneous configurations and diverse CSI distributions, which significantly lowers the overhead of data collection and model fine-tuning when adapting to new scenarios.

\subsection{Representative Roadmaps}
Depending on whether task-specific designs for physical-layer applications are incorporated into the network architecture and the pretraining scheme, we outline two implementation roadmaps for wireless foundation models, as shown in Fig. \ref{paradigm-fig}.
\subsubsection{Building universal wireless representations} 
The first roadmap targets a specific wireless modality by learning a universal wireless representation via self-supervised pretraining, and then applying it to downstream tasks associated with that modality through fine-tuning. 
For example, for two-dimensional spatial–frequency CSI, LWM \cite{LWM} learns general representations through masked channel modeling (MCM) self-supervised learning and is fine-tuned for multiple downstream wireless tasks, such as beam prediction, demonstrating strong few-shot performance. 
In addition, LWLM \cite{LWLM} proposes a hybrid self-supervised pretraining strategy for CSI that combines masking and contrastive learning, yielding substantial improvements in both accuracy and generalization across four downstream wireless localization tasks.

\subsubsection{Enabling wireless task-aware capability}
Nevertheless, since the first roadmap is not designed for dedicated wireless tasks, it still leaves room for improvement on specific physical-layer tasks, particularly under challenging settings. 
Moreover, it is largely limited to fine-tuning and lacks zero-shot generalization, which can reduce but not fundamentally eliminate the cost of task-specific adaptation. 
In fact, for scenario-agnostic wireless tasks (e.g., channel prediction and estimation), it is promising to achieve high-performance zero-shot generalization through wireless task-aware network design and pretraining strategies.
Therefore, building on roadmap 1, the second roadmap aims to endow the model with strong zero-shot inference capabilities across a wide range of scenario-agnostic wireless tasks.
The WiFo family is a representative example of this roadmap, including WiFo \cite{wifo}, WiFo-CF, and WiFo-2 \cite{FM_IWC}.
For instance, WiFo-2 introduces a masked denoising autoencoder architecture and adopts a hybrid suite of self-supervised pretraining objectives, including time- and frequency-domain masked reconstruction for channel prediction as well as interpolation-based denoising for channel estimation. WiFo-2 is the first to achieve unified high-performance zero-shot channel estimation and channel prediction, and it demonstrates state-of-the-art results across eight downstream wireless tasks.

\subsection{Design Principles under Wireless Constraints}
For paradigm 2, the design principles under wireless-specific constraints can be summarized as follows.
\begin{itemize}
    \item Compared with paradigm 1, paradigm 2 typically has a smaller parameter count and lower inference overhead, substantially reducing the storage and computational demands on communication devices. It is also better suited to highly latency-sensitive physical-layer tasks. Moreover, by virtue of its native wireless awareness, paradigm 2 produces more deterministic outputs and delivers stronger inference performance, thereby meeting the reliability requirements of communication systems.
    \item Compared with roadmap 1, roadmap 2 offers superior zero-shot inference performance, making direct deployment in real-world settings feasible. This can fundamentally remove the need for data collection and model fine-tuning, thereby substantially reducing the overhead of online model updates on communication devices. Moreover, because many existing base stations are equipped only with SoCs or FPGAs, online training is often impractical, further underscoring the attractiveness of roadmap 2.
    \item For zero-shot inference, a key challenge is how to guarantee the accuracy of the inferred results. Encouragingly, WiFo-2 achieves reliable accuracy estimation for channel reconstruction through a confidence-enhanced pretraining procedure. This means that, beyond obtaining predictions, we can also assess their expected fidelity. Such capability is crucial for guiding model switching and pilot-density selection, thereby improving the reliability of wireless transmissions.
\end{itemize}
\subsection{Key Features}
The key characteristics of paradigm 2 and a comparison between the two roadmaps are summarized as follows.
\begin{itemize}
    \item Compared with approaches that rely on general-purpose foundation models, paradigm 2 provides native capabilities tailored to wireless tasks. 
    First, consistent with scaling laws, large-scale wireless foundation models can exhibit strong wireless modeling capacity, effectively handling challenging wireless tasks. 
    Second, by learning universal wireless representations from heterogeneous data distributions, they achieve strong few-shot and zero-shot performance under new system configurations and in previously unseen scenarios. 
    Finally, wireless foundation models enable a single model to serve multiple tasks, substantially reducing the number of models required.
    \item Roadmap 1 is well-suited to relatively simple wireless tasks but lacks task-specific enhancement mechanisms, leaving its ability to handle challenging wireless tasks inherently limited. 
    Roadmap 2 addresses this limitation and has led to the emergence of strong capabilities for multiple wireless tasks. 
    For example, by pretraining on large-scale heterogeneous CSI datasets, WiFo-2 achieves superior zero-shot channel estimation and prediction performance, substantially surpassing the full-shot performance of task-specific models.
\end{itemize}

In summary, both paradigms above provide promising schemes for realizing LWFM. A systematic comparison of the two paradigms is presented in Table \ref{paradigm-table}.
\begin{table*}[!ht]
\centering
\footnotesize
\caption{Comparison between the proposed two paradigms of LWFM.}
\label{paradigm-table}
\setlength{\tabcolsep}{4pt}
\renewcommand{\arraystretch}{1.15}

\begin{tabular*}{\textwidth}{@{\extracolsep{\fill}}|P{1.45cm}|P{1.9cm}|P{6.25cm}|P{6.2cm}|}
\hline
\multicolumn{2}{|c|}{} & \multicolumn{1}{c|}{\textbf{Paradigm 1}} & \multicolumn{1}{c|}{\textbf{Paradigm 2}} \\ \hline

\multicolumn{2}{|P{4.8cm}|}{Source of capability}
& General-purpose knowledge transfer
& Large-scale pretraining for wireless domain \\ \hline

\multirow{2}{*}{Features} & Strengths
& Standing on the shoulders of giants
& Emerging native wireless capabilities under scaling laws \\ \cline{2-4}

& Weaknesses
& Constrained by the modality scope of general-purpose foundation models or performance-limited by the domain gap
& Requires the cost-intensive curation of large-scale datasets and subsequent pretraining \\ \hline

\multirow{3}{*}{Capability} & Problem-solving
& Suitable for simple wireless tasks
& Excels at handling challenging wireless tasks \\ \cline{2-4}

& Generalization
& Satisfactory performance under mild distribution shift with few-shot fine-tuning
& Significantly reduces or even eliminates fine-tuning \\ \cline{2-4}

& Universality
& Constrained by the limited variety of task types and system configurations
& Simultaneously adapts to heterogeneous system configurations and a wide range of wireless tasks \\ \hline

\multirow{2}{*}{Requirements} & Hardware-level
& High computational and storage requirements, suitable for the BS deployment
& Significantly reduces storage and computational requirements, enabling potential deployment on resource-constrained edge devices \\ \cline{2-4}

& System-level
& Well-suited to latency-tolerant wireless tasks
& Well-suited to low-latency and high-reliability wireless tasks \\ \hline
\end{tabular*}
\end{table*}

\section{Case Studies: LWFM-Empowered AI-Native Wireless System Design}
\begin{figure*}[!ht]
    \centering
    \includegraphics[width=1\linewidth]{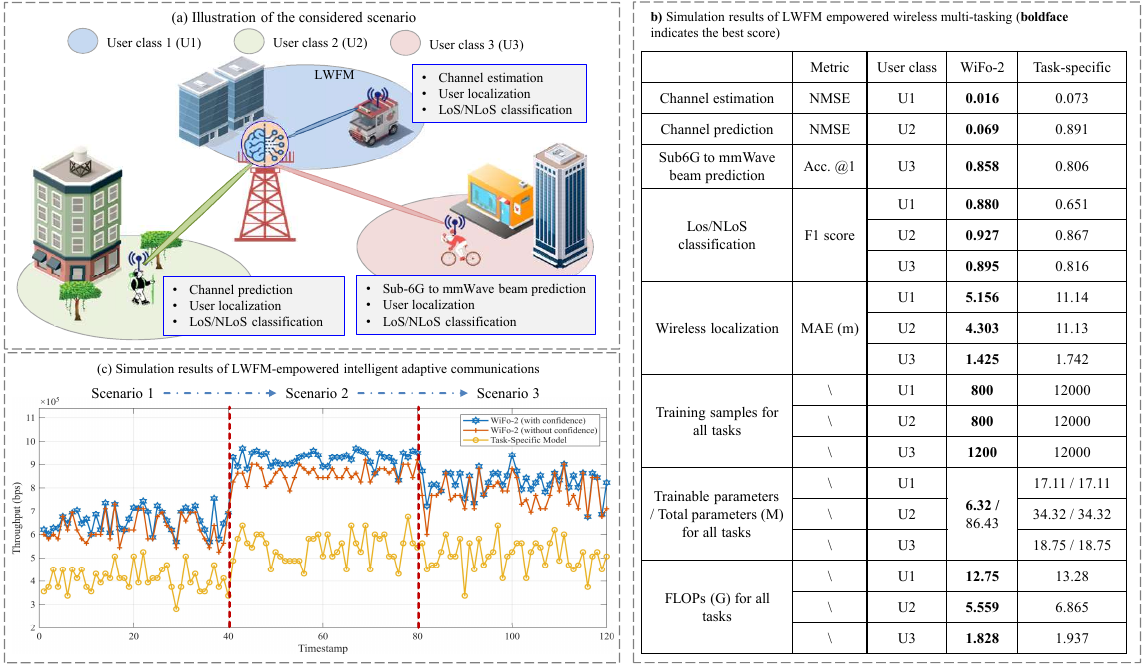}
    \caption{An illustration of the case study for LWFM-empowered AI-native communication system design. (a) illustrates the considered wireless multi-tasking scenario, where the BS serves multiple classes of users. (b) presents the simulation results of LWFM-enabled wireless multi-tasking. (c) presents the simulation results of LWFM-empowered intelligent adaptive communications.}
    \label{case study}
\end{figure*}
This section presents two case studies that demonstrate the advantages of LWFM-empowered AI-native wireless systems through wireless multi-tasking and intelligent adaptive communications.

\subsection{LWFM-Empowered Wireless Multi-Tasking}
For the first experiment, we consider the application of LWFMs in wireless multi-task scenarios.
As shown in Fig.~\ref{case study}~(a), we consider an uplink scenario with a single BS and multiple users, where users exhibit heterogeneous configurations and service requirements.
The BS performs multiple wireless communication and sensing tasks based on the users’ uplink CSI. 
Specifically, there are three classes of users, each associated with a distinct wireless communication task: 
\begin{itemize}
  \item User class 1 (\textbf{channel prediction}): It predicts future CSI according to historical CSI and adopts the normalized mean squared error (NMSE) as the accuracy metric.
  \item User class 2 (\textbf{channel estimation}): It estimates the complete CSI from coarse estimated CSI at sparse pilot locations and adopts NMSE as the accuracy metric.
  \item User class 3 (\textbf{sub-6G to mmWave beam prediction}): It predicts the best mmWave beam codeword index via CSI at sub-6 GHz and adopts top-1 accuracy as the metric.
\end{itemize}
In addition, the BS performs two sensing tasks for each user class as follows.
\begin{itemize}
    \item \textbf{User localization}: 
    It estimates the user’s 2D coordinates relative to the BS via CSI and adopts mean absolute error (MAE) as the accuracy metric.
    \item \textbf{LoS/NLoS classification}: It determines whether a LoS path exists given CSI, and adopts the F1 score as the classification accuracy metric.
\end{itemize}

Notably, all communication and sensing tasks take CSI as the input.
Therefore, we deploy an SOTA LWFM model, WiFo-2, at the BS to solve all tasks with a single model. 
For the channel prediction and estimation task, WiFo-2 performs zero-shot inference without fine-tuning. 
For the remaining tasks, we freeze the WiFo-2 encoder and fine-tune lightweight task-specific output heads attached to it.
Compared with designing a separate model for each task, an LWFM can substantially reduce the number of models and the deployment overhead.

\subsection{LWFM-Empowered Intelligent Adaptive Communications}
In this experiment, we aim to evaluate the LWFM’s ability to adapt to highly dynamic scenarios.
In realistic deployments, channel conditions vary over time due to mobility and environmental dynamics, making a fixed pilot configuration suboptimal. 
Pilot density induces a trade-off between pilot overhead and channel estimation accuracy, which motivates the need for online, throughput-oriented pilot adaptation.
Leveraging WiFo-2's confidence output, the BS can predict the channel estimation accuracy under candidate pilot densities and, accordingly, infer the reliability of the resulting data transmission. 
It then selects the pilot density that best balances pilot overhead against retransmission risk for the current channel condition. 
This design adapts pilot density to instantaneous channel conditions, improving robustness and throughput under dynamics.
Moreover, it supports throughput-oriented pilot selection without additional dedicated air-interface measurements, thereby reducing signaling overhead and enhancing real-time responsiveness.

\subsection{Simulation Results}
We generate datasets using QuaDRiGa to evaluate the proposed LWFM empowered wireless multi-tasking design in which users in each class have heterogeneous antenna and subcarrier configurations.
We also present task-specific baselines consistent with \cite{FM_IWC}, which train and deploy dedicated models for each user and wireless task.
Fig. \ref{case study} (b) shows the performance across different tasks, the total number of training samples aggregated over all tasks, the trainable and total parameters, and the overall FLOPs of each user class.
Leveraging its strong generalization capability, WiFo-2 improves the performance on wireless communication and sensing tasks by \textbf{58.9\%} and \textbf{30.8\%}, respectively, while using only \textbf{7.8\%} of the training samples and \textbf{9.0\%} trainable parameters required by task-specific AI models.
Meanwhile, WiFo-2 achieves total inference FLOPs comparable to the baseline schemes and does not incur a significant increase in computational overhead.
These gains validate the strong zero-shot capability and transferable channel representations of LWFM, which are of great significance for reducing the model deployment overhead in wireless multi-tasking systems.

We further simulate three time-varying scenarios to evaluate the performance of LWFM-empowered intelligent adaptive communications. 
We consider a harsh communication setting with an SNR of -3 dB, where the scenario is updated every 40 time slots and sequentially switches among UMa-NLOS, UMi-LOS, and UMa-LOS.
We consider five pilot densities, where the pilot spacing in the time domain is fixed at 4, and the pilot spacing in the frequency domain is set to 1, 2, 4, 12, and 24 subcarriers, respectively. 
WiFo-2 selects the optimal pilot density based on the throughput estimated from its confidence outputs, whereas the two baseline schemes always use the third pilot density.
WiFo-2 performs zero-shot inference, whereas the task-specific AI model is trained with 4,000 samples in scenario 1.
As shown in Fig. \ref{case study} (c), confidence-enhanced WiFo-2 consistently achieves the optimal throughput among the compared methods.
Specifically, compared with the variant without confidence outputs and the task-specific AI models, WiFo-2 improves the average throughput by \textbf{6.7\%} and \textbf{64.9\%}, respectively.
This is partly attributable to WiFo-2’s strong zero-shot channel estimation performance under harsh channel conditions, and partly to its accurate confidence outputs, which enable adaptive optimization of pilot density selection.
These results highlight the potential of confidence-aware LWFM inference to support real-time decision-making and to serve as a building block toward AI-native cross-layer optimization.

\section{Discussion and Open Issues}
\begin{figure}[!t]
    \centering
    \includegraphics[width=1\linewidth]{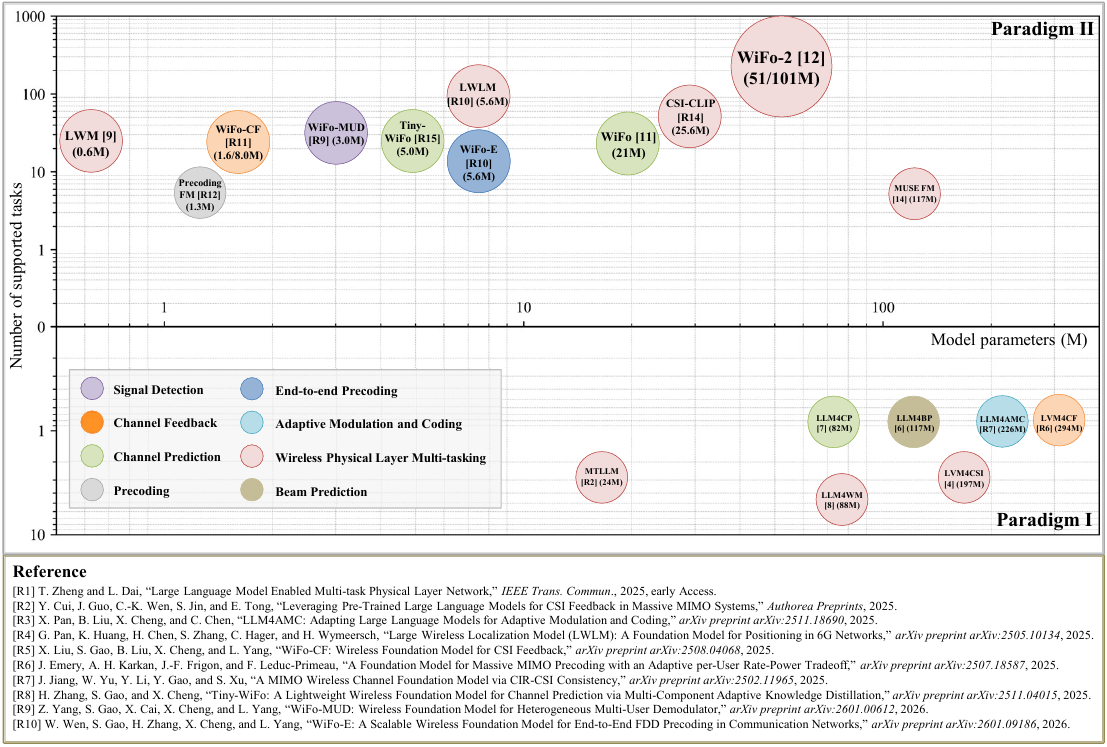}
    \caption{An illustration of the proposed two paradigms for LWFM.}
    \label{large}
\end{figure}
\begin{figure*}[!t]
    \centering
    \includegraphics[width=1\linewidth]{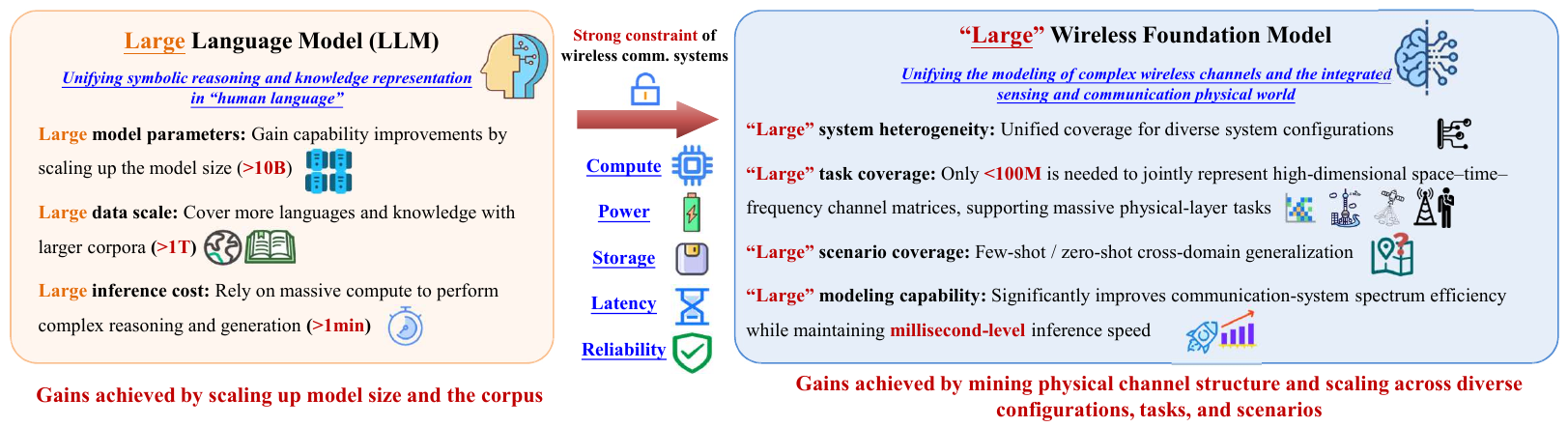}
    \caption{A comparison of the concept of “large” in LLMs and LWFMs.}
    \label{LLM-LWFM}
\end{figure*}
In this section, we first present a comprehensive comparison of existing LWFM studies and then draw a contrast with LLMs to clarify what “large” means in LWFM. 
Finally, we outline several directions for future research.
\subsection{Powerful but Limited Scale}
As shown in Fig. \ref{large}, we provide a multidimensional comparison of representative studies of the two paradigms within a single plot. 
The x-axis reports the number of model parameters to reflect deployment cost. 
WiFo-2 and WiFo-CF adopt an MoE architecture, so we report both the activated and total parameter counts. The y-axis measures the number of supported tasks, defined as the product of wireless task categories and pretraining dataset categories, where datasets with different configurations or distributions are treated as distinct categories.
This metric captures an LWFM’s ability to handle multiple tasks under heterogeneous system configurations and diverse data distributions. 
Most existing LWFMs have fewer than 300~M parameters, and wireless foundation models are typically below 100~M, in sharp contrast to billion-parameter LLMs and better aligned with wireless hardware constraints. Despite their modest scale, LWFMs often support hundreds of tasks, highlighting strong versatility across tasks and settings.
In addition, paradigm 2 supports more tasks with fewer parameters than Paradigm 1, reflecting stronger generality. 
A comparison of the notion of “large” in LLMs and LWFMs is shown in Fig.~\ref{LLM-LWFM}.
Unlike LLMs, “large” in LWFMs refers to capability rather than sheer scale, encompassing broad system heterogeneity, extensive task coverage, wide scenario generalization, and strong inference and modeling capabilities.

\subsection{Future Studies}
Existing efforts on LWFMs remain at an early stage, and building AI-native 6G wireless networks is still a long-term endeavor. 
Looking ahead, the following issues are pressing to bridge the gap between theory and real-world deployment.
\subsubsection{Massive and high-quality dataset construction}
Datasets \cite{dataset} are the cornerstone of LWFM, and their scale and quality ultimately set the upper bound on model performance. 
However, unlike LLMs and LVMs, which can draw on vast collections of text and images from the Internet, constructing large-scale, high-quality wireless datasets faces distinctive challenges, including high costs for data acquisition and processing. 
Existing open-source wireless datasets are typically small in scale and limited in configuration, which constrains large-scale pretraining and hampers fair comparisons across models. 
To fundamentally strengthen the capabilities of LWFMs, it is essential to develop comprehensive wireless datasets, which integrate diverse sources (e.g., real-world measurements, statistical channel models and ray tracing), span a broad range of frequency bands (from sub-6 GHz to terahertz), cover diverse environments (including terrestrial, aerial, space and maritime settings, as well as unmanned aerial vehicle (UAV) scenarios), and encompass varied antenna configurations.
\subsubsection{Empowerment via multimodal information}
In mobile communication networks, rich multimodal information is crucial for reducing pilot overhead and improving link stability. 
However, existing LWFM studies have primarily focused on the CSI modality and have not fully exploited other abundant, heterogeneous multimodal sources \cite{MUSEFM} available within the network, including map information, reference and sounding signals (e.g., demodulation reference signal (DM-RS), sounding reference signal (SRS), and CSI reference signal (CSI-RS)), and multimodal sensing data (e.g., RGB images, LiDAR point clouds, and millimetre-wave radar measurements). 
It is therefore necessary to develop multimodality-enhanced LWFMs, establish interpretable schemes for multimodal information utilization, and ensure strong robustness to flexibly handle missing modalities.
\subsubsection{Extension of physical-layer multi-tasking}
Developing a universal LWFM that enables a single model to support all physical-layer modules represents the ultimate vision of this line of research. 
Given the central role of CSI in wireless systems, previous studies have focused on CSI modality and have shown that one model can benefit multiple CSI-related tasks.
However, the overall coverage of wireless tasks remains to be improved.
On the one hand, it is important to pursue a unified foundation model that can handle the full spectrum of CSI-centric tasks, including less-explored problems such as channel compression and feedback, multi-user precoding, and cross-band channel extrapolation. 
On the other hand, it is equally necessary to extend LWFMs beyond CSI to additional wireless tasks and modalities, such as signal detection, channel decoding, and constellation design.
\subsubsection{Efficient network design and inference acceleration}
As emphasized in this paper, LWFM design is constrained by hardware limitations such as computation and memory, and is also required to satisfy wireless-system constraints, including stringent inference latency. 
Existing studies have reduced inference overhead by adopting efficient architectures such as sparse MoE and by leveraging techniques like knowledge distillation, yet meeting the demands of practical deployment remains challenging. 
Looking ahead, one direction is to explore alternative lightweight architectures, such as replacing transformers with state-space models and developing more efficient attention mechanisms. 
Another direction is to pursue acceleration techniques under wireless inference constraints, including model pruning and low-bit quantization, while also designing hardware-aware acceleration solutions tailored to communication devices.
\subsubsection{Edge–device collaborative intelligence}
Unlike LLMs, which are typically deployed on a single device, wireless communication systems involve multiple devices, so practical LWFM deployment must account for coordination between the BS and end devices. 
Existing LWFMs are not specifically designed for deployment at the edge or on devices, nor do they consider dual-ended or even distributed deployment. 
It is therefore necessary to tailor LWFMs to the distinct characteristics of the BS and the user equipment. 
In addition, further research is needed on cross-device coordination and joint optimization of LWFMs across different communication nodes to improve overall spectral efficiency.

\section{Conclusion}

In this work, we introduce the concept of LWFMs, which empower wireless communication systems via foundation models under wireless constraints. 
We articulate the motivation for bringing foundation models into wireless systems and examine the key design challenges that arise from stringent hardware limitations and system-level constraints. 
To address these challenges, we outline two complementary paradigms for realizing LWFMs. One leverages general-purpose foundation models, while the other builds wireless foundation models tailored to the wireless domain. 
Building on existing efforts, we summarize two representative roadmaps for each paradigm and distill the corresponding design principles under wireless constraints. 
We further present two case studies of LWFM-empowered wireless systems, which demonstrate clear advantages over task-specific AI models in both performance and generalization for wireless multi-tasking and intelligent adaptive communications. 
These results suggest that “large” in LWFMs reflects capability rather than sheer model size. 
Finally, we highlight several open issues for further investigation.

\end{document}